# $^{16}$O+$^{16}$O molecular structure and superdeformation in $^{32}$S

Shigeo Ohkubo

Department of Applied Science and Environment, Kochi Women's University, Kochi 780-8515, Japan

**Abstract.** The molecular structure with the $^{16}$O+$^{16}$O configuration in $^{32}$S and the superdeformed structure in $^{32}$S are discussed from the viewpoint of the cluster model.

Almost two hundreds superdeformed rotational bands have been observed in a wide range of mass regions ($A = 60, 80, 130, 150, 190$) [1]. Recently superdeformation has been observed even in the $sd$-shell region such as $^{38}$Ar and $^{40}$Ca [2,3]. The superdeformed band in $^{32}$S, which is a doubly magic nucleus in the deformed shell model, is very interesting because it gives "the missing link between the known region of superdeformed nuclei around $^{60}$Zn and the cluster-like structures in lighter nuclei, like in $^{20}$Ne and $^{8}$Be " [1]. Yamagami and Matsuyanagi [4] calculated the superdeformed band structure in $^{32}$S in the symmetry-unrestricted cranked Hartree-Fock method using the Skyrme force.

On the other hand, as for the cluster structure in nuclei, the band structure with $\alpha$-cluster configuration has been clearly established for $4N$-nuclei such as $^{8}$Be, $^{20}$Ne and $^{44}$Ti [5]. In order to link the superdeformed structure with the cluster structure in $^{32}$S, it is important to make clear the $^{16}$O + $^{16}$O clustering aspects in $^{32}$S. The $^{16}$O + $^{16}$O structure in $^{32}$S, which is an analog of the $\alpha+\alpha$ structure in $^{8}$Be, has been most extensively studied from the viewpoint of the molecular structure. However, although many states with the $^{16}$O + $^{16}$O molecular structure have been observed, the band structure with the $^{16}$O + $^{16}$O configuration in $^{32}$S has not been established.

We show that the $^{16}$O + $^{16}$O cluster bands exist in $^{32}$S from a different point of view, that is, rainbow scattering. From a systematic analysis of rainbow scattering the interaction potential for the $^{16}$O + $^{16}$O system can be well determined up to very internal region. The potential with a folding-type diffuse surface can describe the elastic scattering angular distributions in a wide range of energies from $E_L=$ 1120 MeV to 75 MeV. It is also shown that this potential can describe the low energy $^{16}O + ^{16}O$ elastic scattering between $E_L = 63$ MeV to 25 MeV, which means that the potential is useful for the study of the $^{16}O + ^{16}O$ structure of $^{32}S$.

In order to reveal the resonant and bound structures of the $^{16}O + ^{16}O$ system, we have solved the complex scaling equation with the real part of the optical potential for the $^{16}O + ^{16}O$ system. We have obtained the rotational bands with the $^{16}O + ^{16}O$ configuration. The first Pauli-allowed band with $N = 24$ appears about 7.5 MeV below the threshold, that is 9 MeV in excitation energy from the ground state of $^{32}S$. The second $N = 26$ band starts at about 3 MeV above the threshold. The $N = 28$ band starts at about 10 MeV above the threshold. The centroids of the observed molecular states in this region correspond well to the calculated $N = 28$ band. Many molecular resonances observed in experiment may be fragmented from the $N = 28$ band. There are no experimental candidates for the $N = 26$ band, which is inevitably predicted in-between the $N = 24$ and $N = 28$ bands. Therefore this band states should be searched for in experiment.

The band head of the superdeformed rotational band by Yamagami and Matsuyanagi [4] appears at about 12 MeV with the standard SIII force, which is about 4 MeV higher than the present calculation. The authors point out that the band head energy becomes about 9 MeV if the rotational zero-point corrections are taken into account [4]. Thus the superdeformed band well corresponds to our $N = 24$ cluster band in excitation energy. Also the structure change of the band occurs at $J = 24$ in Ref.[4], which is consistent with our result that due to the Pauli principle the maximum $J$ of the lowest band is 24. The rotational constant $k$ of the superdeformed band estimated from $J = 10 - 22$ in Ref.[4] is about 50-55 keV, which is consistent with our calculated $k = 52$ keV. Thus the superdeformed rotational band in $^{32}S$ corresponds well to the lowest $N = 24$ cluster band with the $^{16}O + ^{16}O$ configuration proposed in Ref.[6]. The calculated $B(E2)$ values of the $N = 24$ band is enhanced very much.

The author thanks Prof. K. Matsuyanagi for useful conversations on superdeformation.

a. E-mail: shigeo@cc.kochi-wu.ac.jp